\documentclass[prd,twocolumn,nopacs,floatfix,amsmath,nofootinbib,amssymb,preprintnumbers]{revtex4-2}
\usepackage{graphicx,color,dcolumn,booktabs,bm}
\usepackage{txfonts}
\usepackage{overpic}
\usepackage{amssymb}
\usepackage{indentfirst}
\usepackage{feynmf}   
\usepackage{slashed}  
\usepackage{cases}
\usepackage{color}
\usepackage{multirow}
\usepackage{slashed}
\usepackage{epstopdf}
\usepackage{longtable}
\usepackage{ulem}

\usepackage{graphicx,color,dcolumn,booktabs,bm}
\usepackage[colorlinks,
            citecolor=blue,
            anchorcolor=red,
            menucolor=red,
            linkcolor=red,
            filecolor=red,
            runcolor=red,
            urlcolor=blue,
            frenchlinks=red]{hyperref}

\graphicspath{{Figures/}} %
\allowdisplaybreaks

\begin{document}

\title{Construction of Isoscalar $2^{-+}$ Mesonic States Inspired by the $X(2600)$ Discovery}

\author{Li-Ming Wang$^{1,3}$}\email{lmwang@ysu.edu.cn}
\author{Wen-Xin Tian$^{1}$}
\author{Ting-Yan Li$^{2,3,4}$}
\email{lity2023@lzu.edu.cn}
\author{Cheng-Xi Liu$^{2,3,4}$}
\email{liuchx2023@lzu.edu.cn}
\author{Xiang Liu$^{2,3,4}$\footnote{Corresponding author}}\email{xiangliu@lzu.edu.cn}
 \affiliation{
 $^1$Key Laboratory for Microstructural Material Physics of Hebei Province, School of Science, Yanshan University, Qinhuangdao 066004, China\\
$^2$School of Physical Science and Technology, Lanzhou University, Lanzhou 730000, China\\
$^3$Lanzhou Center for Theoretical Physics, Key Laboratory of Theoretical Physics of Gansu Province and Frontiers Science Center for Rare Isotopes, Lanzhou University, Lanzhou 730000, China\\
$^4$Research Center for Hadron and CSR Physics, Lanzhou University and Institute of Modern Physics of CAS, Lanzhou 730000, China}

\begin{abstract}

The BESIII observation of the $X(2600)$ suggests that one possible assignment for its spin-parity quantum numbers is $J^{PC}=2^{-+}$. This finding motivates us to systematically revisit isoscalar pseudotensor mesonic states through a comprehensive spectroscopic study. In this work, we employ a potential model to calculate the mass spectrum of isoscalar pseudotensor meson states with varying radial quantum numbers. Additionally, we analyze their strong decay behaviors, as allowed by the Okubo-Zweig-Iizuka (OZI) rule, using the quark pair creation model. Our findings yield valuable information about their resonant parameters and partial decay widths, which can aid in identifying the observed isoscalar pseudotensor states when combined with current experimental data. Importantly, we also predict several radial excitations of isoscalar pseudotensor mesons that remain unobserved experimentally. Furthermore, we exclude the $X(2600)$ from being classified as an isoscalar pseudotensor meson, which is critical for elucidating the properties of the $X(2600)$.

\end{abstract}
\date{\today}
\maketitle

\section{Introduction}\label{sec1}

Recently, the BESIII Collaboration announced the observation of the $X(2600)$ state in the $\eta^\prime \pi^+\pi^-$ invariant mass spectrum of $J/\psi \to \gamma \eta^\prime \pi^+\pi^-$ \cite{BESIIICollaboration:2022kwh}, with a statistical significance exceeding $20\sigma$. The $X(2600)$ has a measured mass of $2617.8\pm2.1^{+18.2}_{-1.9}$ MeV and a width of $\Gamma=200\pm8^{+20}_{-17}$ MeV. The spin-parity quantum numbers $J^{PC}$ of the $X(2600)$ are yet to be determined, and from its decay mode, it could either be $0^{-+}$ or $2^{-+}$ \cite{BESIIICollaboration:2022kwh}. In Ref. \cite{Wang:2020due}, the $X(2600)$ is proposed to belong to the isoscalar pseudoscalar meson family as an $\eta(6S)$ state. This classification aligns with earlier studies of pseudoscalar mesonic states presented in Refs. \cite{Yu:2011ta,Wang:2017iai}. Currently, the possibility of the $X(2600)$ being a isoscalar pseudotensor meson with $J^{PC}=2^{-+}$ warrants thorough examination through a realistic study, which serves as the motivation for the present work.

We might as well begin by examining {\it The Review of Particle Physics} (RPP) published by the Particle Data Group \cite{ParticleDataGroup:2024cfk} to gather information on the observed isoscalar pseudotensor states. The summary table in the RPP includes the $\eta_2(1645)$ and $\eta_2(1870)$, while the $\eta_2(2030)$ and $\eta_2(2250)$ are listed as {\it further states} \cite{ParticleDataGroup:2024cfk}. Including the previously discussed $X(2600)$ \cite{BESIIICollaboration:2022kwh}, there are five isoscalar pseudotensor states awaiting classification, which is the main task of this work.

In fact, a spectroscopic study of isoscalar pseudotensor mesons is an effective approach to elucidate the properties of these five isoscalar pseudotensor states and to construct a comprehensive isoscalar pseudotensor meson family. We aim to obtain the mass spectrum of isoscalar pseudotensor mesons, which can be calculated using the potential model known as the Godfrey-Isgur model \cite{Godfrey:1985xj}. Additionally, the two-body OZI-allowed strong decays can provide crucial information about the partial and total decay widths, utilizing the quark pair creation (QPC) model \cite{Micu:1968mk, LeYaouanc:1973ldf, LeYaouanc:1977gm, LeYaouanc:1977fsz,Blundell:1996as} in this investigation. By combining these findings with the experimental data for the observed states, we can identify candidates for the isoscalar pseudotensor meson family. This study is not only focused on elucidating the properties of the observed isoscalar pseudotensor states but also predicts several isoscalar pseudotensor mesons that await further experimental investigation. Importantly, we exclude the possibility of the observed $X(2600)$, as 
suggested by BESIII, being part of the isoscalar pseudotensor meson family, which is a significant finding for clarifying the nature of the $X(2600)$.

This paper is organized as follows. Following the introduction in Sec. \ref{sec1}, we present a spectroscopic analysis of isoscalar pseudotensor mesons, including their mass spectrum and two-body OZI-allowed decays. Additionally, we analyze the observed isoscalar pseudotensor states and predict several missing states (see Sec. \ref{sec2}). The paper concludes with a brief summary in Sec. \ref{sec3}.

\section{Spectroscopic analysis}
\label{sec2} 
\subsection{Mass spectrum calculation}

In calculating the mass spectrum of the focus isoscalar pseudotensor meson, we employ the Godfrey-Isgur (GI) model, which has demonstrated significant effectiveness in describing the mass spectra of mesons, particularly for low-lying states \cite{Capstick:1986ter,Barnes:2005pb}. For higher excited states, it is crucial to introduce a screening potential to account for unquenching effects \cite{Song:2015nia, Song:2015fha, Pang:2017dlw, Wang:2018rjg, Wang:2019mhs, Wang:2020prx, Wang:2021abg}. Additionally, we will briefly outline some key details of the GI model.

The interaction between a quark and an antiquark is represented by a potential model Hamiltonian, which encompasses both kinetic energy terms and an effective potential term
\begin{equation}\label{Hamtn}
  \tilde{H}=\sqrt{m_1^2+\mathbf{p}^2}+\sqrt{m_2^2+\mathbf{p}^2}+\tilde{V}_{\mathrm{eff}}(\mathbf{p,r}),
\end{equation}
where $m_1$ and $m_2$ represent the masses of the quark and antiquark, respectively. The effective potential, $\tilde{V}_{\mathrm{eff}}$, consists of two main components: a short-range one-gluon-exchange interaction, expressed as $\gamma^{\mu}\otimes\gamma_{\mu}$, and a linear confinement interaction, represented as $1\otimes1$. The meaning of the tilde symbol will be clarified later.

In the nonrelativistic limit, the effective potential has a familiar format \cite{Godfrey:1985xj}
\begin{eqnarray}
V_{\mathrm{eff}}(r)=H^{\mathrm{conf}}+H^{\mathrm{hyp}}+H^{\mathrm{so}}\label{1}
\end{eqnarray}
with
\begin{align}
 H^{\mathrm{conf}}&=\Big[-\frac{3}{4}(c+br)+\frac{\alpha_s(r)}{r}\Big](\bm{F}_1\cdot\bm{F}_2)\nonumber\\ &=S(r)+G(r)\label{3}\\
H^{\mathrm{hyp}}&=-\frac{\alpha_s(r)}{m_{1}m_{2}}\Bigg[\frac{8\pi}{3}\bm{S}_1\cdot\bm{S}_2\delta^3 (\bm r) +\frac{1}{r^3}\Big(\frac{3\bm{S}_1\cdot\bm r \bm{S}_2\cdot\bm r}{r^2} \nonumber  \\ \label{3.1}
&\quad -\bm{S}_1\cdot\bm{S}_2\Big)\Bigg] (\bm{F}_1\cdot\bm{F}_2),  \\
H^{\mathrm{so}}=&H^{\mathrm{so(cm)}}+H^{\mathrm{so(tp)}},  \label{3.2}
\end{align}
where $H^{\mathrm{conf}}$ includes the spin-independent linear confinement piece $S(r)$ and Coulomb-like potential from one-gluon-exchange $G(r)$. $H^{\mathrm{hyp}}$ denotes the color-hyperfine interaction consisting of tensor and contact terms. 
 $H^{\mathrm{SO}}$ is the spin-orbit interaction with
\begin{eqnarray}
H^{\mathrm{so(cm)}}=\frac{-\alpha_s(r)}{r^3}\left(\frac{1}{m_{1}}+\frac{1}{m_{2}}\right)\left(\frac{\bm{S}_1}{m_{1}}+\frac{\bm{S}_2}{m_{2}}\right)
\cdot
\bm{L}(\bm{F}_1\cdot\bm{F}_2),
\end{eqnarray}
which is caused by one-gluon-exchange, while
\begin{eqnarray}
H^{\mathrm{so(tp)}}=-\frac{1}{2r}\frac{\partial H^{\mathrm{conf}}}{\partial
r}\Bigg(\frac{\bm{S}_1}{m^2_{1}}+\frac{\bm{S}_2}{m^2_{2}}\Bigg)\cdot \bm{L}
\end{eqnarray}
is the Thomas precession term.

In the above equations, $\bm{S}_1$ and $\bm{S}_2$ represent the spins of the quark and antiquark, respectively, while $\bm{L}$ denotes the orbital angular momentum between the two particles. The term $\bm{F}$ is associated with the Gell-Mann matrices, where $\bm{F}_1 = \bm{\lambda}_1/2$ and $\bm{F}_2 = -\bm{\lambda}^*_2/2$. For mesons, the expectation value is given by $\langle\bm{F}_1\cdot\bm{F}_2\rangle = -4/3$.

To account for relativistic effects, which have a significant impact on meson systems, two key modifications are introduced. First, a smearing function is applied to account for nonlocal interactions and the new $\mathbf{r}$ dependence in a meson composed of a quark-antiquark pair ($q\bar{q}$):

\begin{equation}
\rho \left(\mathbf{r}-\mathbf{r'}\right)=\frac{\sigma^3}{\pi ^{3/2}}e^{-\sigma^2\left(\mathbf{r}-\mathbf{r'}\right)^2},
\end{equation}
which is applied to $S(r)$ and $G(r)$ to obtain smeared potentials $\tilde{S}(r)$ and $\tilde{G}(r)$ by
\begin{equation}\label{smear}
\tilde{f}(r)=\int d^3r'\rho(\mathbf{r}-\mathbf{r'})f(r')
\end{equation}
with
\begin{eqnarray}
   \sigma_{12}^2=\sigma_0^2\Bigg[\frac{1}{2}+\frac{1}{2}\left(\frac{4m_1m_2}{(m_1+m_2)^2}\right)^4\Bigg]+
  s^2\left(\frac{2m_1m_2}{m_1+m_2}\right)^2.
\end{eqnarray}

Second, due to relativistic effects, the potential must depend on the center-of-mass of the interacting quarks. To account for this, momentum-dependent factors are introduced, which reduce to unity in the non-relativistic limit. These adjustments are applied as follows:

\begin{equation}
\tilde{G}(r)\to \left(1+\frac{p^2}{E_1E_2}\right)^{1/2}\tilde{G}(r)\left(1 +\frac{p^2}{E_1E_2}\right)^{1/2},
\end{equation}
and
\begin{equation}
  \frac{\tilde{V}_i(r)}{m_1m_2}\to \left(\frac{m_1m_2}{E_1E_2}\right)^{1/2+\epsilon_i} \frac{\tilde{V}_i(r)}{m_1 m_2} \left( \frac{m_1 m_2}{E_1 E_2}\right)^{1/2+\epsilon_i},
\end{equation}
where $\tilde{V}_i(r)$ represents the contact, tensor, vector spin-orbit, and scalar spin-orbit terms, with $\epsilon_i$ being the corresponding modification parameter.

The screening effect is introduced through the transformation: $br+c\rightarrow \frac{b(1-e^{-\mu r})}{\mu}+c$, where $\mu$ is the screening parameter, the specific value of which can be found in Ref. \cite{Pang:2018gcn}. Additionally, the modified confinement potential requires similar relativistic corrections, as outlined in the GI model.
 Then, we further write
\begin{eqnarray}
&&\tilde V^{\mathrm{screen}}(r)\nonumber\\&&= \int d^3 \bm{r}^\prime
\rho (\bm{r-r^\prime})\frac{b(1-e^{-\mu r'})}{\mu}\nonumber\\
&&= \frac{b}{\mu r}\Bigg[r+e^{\frac{\mu^2}{4 \sigma^2}+\mu r}\frac{\mu+2r\sigma^2}{2\sigma^2}\Bigg(\frac{1}{\sqrt{\pi}}
\int_0^{\frac{\mu+2r\sigma^2}{2\sigma}}e^{-x^2}dx-\frac{1}{2}\Bigg) \nonumber\\
&&\quad-e^{\frac{\mu^2}{4 \sigma^2}-\mu r}\frac{\mu-2r\sigma^2}{2\sigma^2}\Bigg(\frac{1}{\sqrt{\pi}}
\int_0^{\frac{\mu-2r\sigma^2}{2\sigma}}e^{-x^2}dx-\frac{1}{2}\Bigg)\Bigg]
.\label{5}
\end{eqnarray}

To calculate the mass spectrum and spatial wave functions of isoscalar pseudotensor mesons, denoted as $\eta_2(nD)$ and $\eta_2^\prime(nD)$\footnote{Here, $\eta_2(nD)$ represents  $n\bar{n}$ ($n=u,d$) state, while $\eta_2^\prime(nD)$ denotes $s\bar{s}$ state.}, one needs to solve the eigenvalues and eigenvectors of the Hamiltonian $\tilde{H}$, as given in Eq. (\ref{Hamtn}). The parameters involved in the GI model are collected in Table \ref{parameters}, which are taken from the Ref. \cite{Wang:2021abg}.
This is typically done using the simple harmonic oscillator (SHO) basis expansion method. 
\ In this approach, the precise wave function $\Phi(\mathbf{r})$, obtained as an eigenvector of the Hamiltonian $\tilde{H}$ ( Eq. (\ref{Hamtn})), is expressed as a linear combination of SHO wave functions $\Psi_{nLM_L}(\mathbf{r})$, which form a complete orthonormal basis set. These wave functions are then used in subsequent calculations of the strong decays of $\eta_2^{(\prime)}$ mesons, specifically for the spatial wave function components of the initial and final states.
The expansion allows us to write:  
\begin{equation}\label{2.10}
\int |\Psi_{nLM_L}({\bf r})|^2{\bf r}^2{\rm d}^3{\bf r}=\int |\Phi({\bf r})|^2{\bf r}^2{\rm d}^3{\bf r},
\end{equation}

The SHO wave functions $\Psi_{nLM_L}(\mathbf{r})$ are explicitly defined in both coordinate space and momentum space as follows:  
\begin{align}\label{SHO}
\Psi_{nLM_L}(\mathbf{r})=R_{nL}(r, \beta)Y_{LM_L}(\Omega_r), \nonumber \\
\Psi_{nLM_L}(\mathbf{p})=R_{nL}(p, \beta)Y_{LM_L}(\Omega_p),  
\end{align}  
where $Y_{LM_L}(\Omega)$ represents the spherical harmonic function, and $R_{nL}(r, \beta)$ is the radial part of the SHO wave function given by:  
\begin{eqnarray}
&R_{nL}(r, \beta) = \beta^{3/2}\sqrt{\frac{2n!}{\Gamma(n+L+3/2)}}(\beta r)^{L} e^{-\frac{r^2 \beta^2}{2}} \nonumber \\
&\quad \times L_{n}^{L+1/2}(\beta^2r^2),\\
&R_{nL}(p, \beta) = \frac{(-1)^n(-i)^L}{\beta^{3/2}} e^{-\frac{p^2}{2 \beta^2}} \sqrt{\frac{2n!}{\Gamma(n+L+3/2)}} {(\frac{p}{\beta})}^{L} \nonumber \\
&\quad \times L_{n}^{L+1/2}(\frac{p^2}{\beta^2}),
\end{eqnarray}  
where $L_{n}^{L+1/2}(x)$ is the associated Laguerre polynomial.  

By expanding $\Phi(\mathbf{r})$ in terms of $\Psi_{nLM_L}(\mathbf{r})$, we can effectively utilize the SHO basis to simplify the computation of observables, such as decay widths. This method ensures that the precise wave function retains all physical corrections introduced by the Hamiltonian $\tilde{H}$, while leveraging the mathematical tractability of the SHO basis.

\begin{table}[htbp]
\caption{The value of these parameters involved in the MGI model \cite{Wang:2021abg}.\label{parameters}}
\renewcommand\arraystretch{1.2}
\begin{tabular*}{86mm}{c@{\extracolsep{\fill}}cccc}
\toprule[1pt]\toprule[1pt]
Parameter                 & Value  & Parameter                 & Value  \\
\toprule[0.8pt]
$m_u$ (GeV)               & 0.22   & $m_d$ (GeV)               & 0.22   \\
$m_s$ (GeV)               & 0.424  & $b$ (GeV$^2$)             & 0.229  \\
$\epsilon_c$              & -0.164 & $\epsilon_{\mathrm{sos}}$ & 0.9728 \\
$\sigma_0$ (GeV)          & 1.8    & $s$ (GeV)                 & 3.88   \\
$\mu$ (GeV)               & 0.081  & $c$ (GeV)                 & -0.30 \\
$\epsilon_{\mathrm{sov}}$ & 0.262  & $\epsilon_t$              & 1.993  \\
\toprule[1pt]
\end{tabular*}\label{para}
\end{table}

\begin{table}[htbp]
\centering
\caption{The comparison of the calculated results and  the experimental data of the masses for the isoscalar pseudotensor mesons. All results are in units of MeV.  \label{mass}}
\renewcommand\arraystretch{1.5}
\begin{tabular*}{80mm}{@{\extracolsep{\fill}}lccc}
\toprule[1.00pt]
\toprule[1.00pt]
  States & This work & Experimental values &\\
\hline
   $\eta_2(1D)$      & 1650   &$1645\pm14\pm15$ \cite{CrystalBarrel:1996bnu} \\
 $\eta_2(2D)$      &   2003 & $2030\pm10\pm15$ \cite{Anisovich:2000mv}\\
 $\eta_2(3D)$      &   2279  &$2267\pm14$  \cite{Anisovich:2000ut}   \\
 $\eta_2(4D)$     &   2498  &$...$    \\
\hline
 $\eta^{\prime}_2(1D)$         & 1882    &$1881\pm32\pm40$ \cite{CrystalBall:1991zkb}\\
 $\eta^{\prime}_2(2D)$           &  2238 & $...$    \\
 $\eta^{\prime}_2(3D)$     & 2520  &$...$   \\
 $\eta^{\prime}_2(4D)$     &  2764 &$...$    \\
\bottomrule[1pt]
\end{tabular*}
\end{table}

In Table \ref{mass}, we present the calculated mass spectrum of isoscalar pseudotensor mesons and compare it with four reported $\eta_2$ states compiled by the RPP. These states include the $\eta_2(1645)$ and $\eta_2(1870)$, as well as the $\eta_2(2030)$ and $\eta_2(2250)$, which are categorized as "further states." Among these, the $\eta_2(1645)$ is considered the ground state of the $\eta_2$ meson, while the $\eta_2(2030)$ and $\eta_2(2250)$ are identified as its first and second radial excitations, respectively. The state $\eta_2(1870)$ is classified as the ground state of the $\eta^{\prime}_2$ meson. Our calculated mass spectrum in Table \ref{mass} supports these assignments.

We predict the masses of the $\eta_2(4D)$, $\eta_2^\prime(2D)$, and $\eta_2^\prime(3D)$ to be 2498 MeV, 2238 MeV, and 2520 MeV, respectively. Notably, the theoretical mass of the $\eta_2^\prime(4D)$ is 2764 MeV, leading us to consider a potential connection between this state and the observed $X(2600)$ by BESIII. However, we must acknowledge the mass difference when interpreting the $X(2600)$ as the $\eta_2^\prime(4D)$ state.

\subsection{Strong decay behavior}\label{subsec2}

The Quark Pair Creation (QPC) model \cite{Micu:1968mk, LeYaouanc:1973ldf, LeYaouanc:1977gm, LeYaouanc:1977fsz,Blundell:1996as}, initially proposed by Micu, has since been widely applied to calculate strong two-body decays allowed by the OZI rule, following further advancements \cite{Wang:2019qyy, Chen:2019ywy, Luo:2023sne, Pang:2017dlw, Wang:2016krl, Li:2020xzs, li:2021hss, Ni:2021pce, Li:2023wgq, Li:2022bre, Wang:2019mhs, Liu:2024xav}. For a decay process $A \to B + C$, the transition matrix is defined as \cite{vanBeveren:1979bd, Blundell:1996as}
\begin{equation}\label{3.6}
\langle BC|\mathcal{T}|A \rangle = \delta ^3(\boldsymbol{{P}_B}+\boldsymbol{{P}_C)}\mathcal{M}^{{M}_{J_{A}}M_{J_{B}}M_{J_{C}}},
\end{equation}
the amplitude
$\mathcal{M}^{M_{J_A}M_{J_B} M_{J_C}}(\mathbf{P})$ can be derived.  $\mathcal{T}$ can be expressed as

\begin{align}\label{3.7}
\mathcal{T}& = -3\gamma \sum_{m}\langle 1m;1~-m|00\rangle\int d {\boldsymbol{p}}_3d {\boldsymbol{p}}_4\delta ^3 ({\boldsymbol{p}}_3+{\boldsymbol{p}}_4) \nonumber \\
 & ~
 \quad\times \mathcal{Y}_{1m}\left(\frac{{\boldsymbol{p}}_3-{\boldsymbol{p}}_4}{2}\right)\chi _{1,-m}^{34}\phi _{0}^{34}
\left(\omega_{0}^{34}\right)_{ij}b_{3i}^{\dag}({\boldsymbol{p}}_3)d_{4j}^{\dag}({\boldsymbol{p}}_4),
\end{align}
This is the transition operator, which describes the creation of a quark-antiquark pair from the vacuum. Here, $b_3^{\dag}(d_{4}^{\dag})$ represents the quark (antiquark) creation operator, while $\chi$, $\phi$, and $\omega$ denote the spin, flavor, and color wave functions, respectively. The parameter $\gamma$ is a dimensionless constant that characterizes the pair production rate from the vacuum. The function $\mathcal{Y}_{\ell m}({p}) = {|{p}|^{\ell}}Y_{\ell m}({p})$ is a solid spherical harmonic. By applying the Jacobi-Wick formula \cite{Jacob:1959at}, $\mathcal{M}^{M_{J_A} M_{J_B} M_{J_C}}(\mathbf{P})$ is transformed into partial wave amplitudes $\mathcal{M}^{JL}$, which can be expressed as follows:
\begin{align}\label{3.8}\begin{split}
\mathcal{M}^{J L}({\boldsymbol{P}})=& \frac{\sqrt{4 \pi(2 L+1)}}{2 J_{A}+1} \sum_{M_{J_{B}} M_{J_{C}}}\left\langle L 0 ; J M_{J_{A}} \mid J_{A} M_{J_{A}}\right\rangle \\
& \times\left\langle J_{B} M_{J_{B}} ; J_{C} M_{J_{C}} \mid J_{A} M_{J_{A}}\right\rangle \mathcal{M}^{M_{J_{A}} M_{J_{B}} M_{J_{C}}}.
\end{split}\end{align}
Then the strong decay partial width for a given decay mode of $A \rightarrow B+C$ reads as
\begin{align}\label{3.9}
\Gamma=\frac{\pi}{4} \frac{|P_E|}{m_{A}^{2}} \sum_{J, L}\left|\mathcal{M}^{J L}({\boldsymbol{P}})\right|^{2},
\end{align}
where $\mathrm{m}_{A}$ denotes the mass of the initial meson $A$. For our calculations, we require the spatial wave functions of the mesons involved, which are numerically obtained using the modified GI model. In this work, we use $\gamma = 6.57$ and $\gamma_s = 6.57/\sqrt{3}$.

In the following four cases, we calculate the OZI-allowed decay behavior of the $\eta_2$ meson family using the QPC model. 

we found that the mixing of flavor wave functions $|n\bar{n}\rangle$ (where $n = u,d$) and $|s\bar{s}\rangle$ cannot be neglected, as supported by several experimental observations. For the $\eta_2(1870)$, experiments have observed its decay into $a_2(1320)\pi$ and $a_0(980)\pi$ (as shown in the RPP \cite{ParticleDataGroup:2024cfk}). These results indicate that the $\eta_2(1870)$ cannot be a pure $s\bar{s}$ state. This is the main reason we introduced the mixing scheme of flavor wave functions $|n\bar{n}\rangle = (|u\bar{u}\rangle + |d\bar{d}\rangle)/\sqrt{2}$ and $|s\bar{s}\rangle$ for all $\eta_2^{(\prime)}$ mesons. A detailed discussion is provided in the following section. The states $\eta_2(nD)$ and $\eta^{\prime}_2(nD)$ satisfy the following mixing relation:
\begin{equation}\label{mixingangleh11P}
\left( \begin{matrix}
	|\eta_2(nD)\rangle \\
	|\eta^{\prime}_2(nD)\rangle \\
\end{matrix}\right) =
\left( \begin{matrix}
	\textrm{$\cos\theta$} & \textrm{$-\sin\theta$} \\
	\textrm{$\sin\theta$} & \textrm{$\cos\theta$} \\
\end{matrix}\right)
\left( \begin{matrix}
	|n\bar{n}\rangle \\
	|s\bar{s}\rangle \\
\end{matrix}\right),
\end{equation}
where $\eta^{\prime}_2(nD)$ is the partner of the $\eta_2(nD)$, and $\theta$ denotes the mixing angle.

\subsubsection{The $n=1$ case}

In Table \ref{1d}, we present the strong decay behavior of the $\eta_2(1D)$ and $\eta^{\prime}_2(1D)$. The results indicate that the total decay width of the $\eta_2(1645)$ decreases with the mixing angle, with the decay width aligning well with the central value of the experimental data (180 MeV \cite{CrystalBarrel:1996bnu}) at $\theta = 30^{\circ}$. According to the PDG, the observed decay channels of the $\eta_2(1645)$ include $a_2(1320) \pi$, $K \bar{K} \pi$, $K^* \bar{K}$, $\eta \pi^+ \pi^-$, and $a_0(980) \pi$. Our findings suggest that $\pi a_2(1320)$ is the dominant decay channel, with $\rho \rho$ making a significant contribution, while the $K K^*$ channel contributes less. The calculated ratio $\Gamma(K K^*) / \Gamma(a_2(1320) \pi) = 0.06$ is in close agreement with the experimental value $\Gamma(K \bar{K} \pi) / \Gamma(a_2(1320) \pi) = 0.07 \pm 0.02 \pm 0.02$ \cite{WA102:1997gkz}. This result supports the hypothesis that $\eta_2$ mesons contain a mixture of flavor components.

\begin{figure}[hptb]
\begin{center}
	\scalebox{1.0}{\includegraphics[width=1.0\columnwidth]{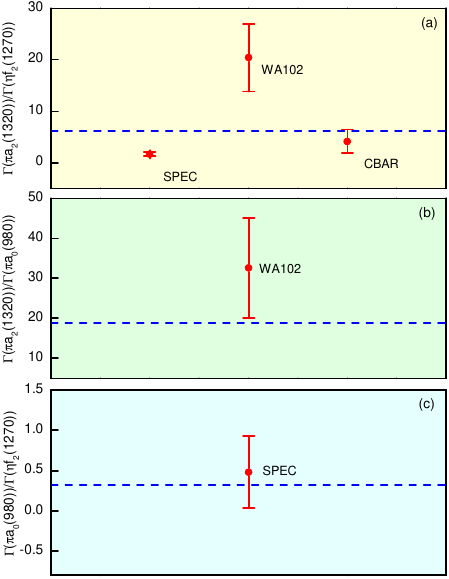}}
	\caption{Comparison of theoretical and experimental values for three decay branching ratios. Panel (a) compares the theoretical and experimental values of $\Gamma(\pi a_2(1320))/\Gamma(\eta f_2(1270))$; panel (b) compares $\Gamma(\pi a_2(1320))/\Gamma(\pi a_0(980))$; and panel (c) compares $\Gamma(\pi a_0(980))/\Gamma(\eta f_2(1270))$. The red dashed lines represent theoretical values, while the purple circles with error bars denote experimental values.}\label{fig1}
\end{center}
\end{figure}

For the decay properties of the $\eta_2(1870)$, we calculated its OZI-allowed decay widths for mixing angles $\theta = 0^{\circ}, 15^{\circ},$ and $30^{\circ}$ (see Table \ref{1d}). The results show that $\pi a_2(1320)$ is the dominant decay channel for $\theta = 15^{\circ}$ and $\theta = 30^{\circ}$. At $\theta = 0^{\circ}$, where the flavor content is purely $|s\bar{s}\rangle$, the $\eta_2(1870)$ decays exclusively into the $K K^*$ and $K^* K^*$ channels, which also contribute significantly at $\theta = 15^{\circ}$ and $\theta = 30^{\circ}$. Given the large experimental uncertainty for this state, our results at $\theta = 30^{\circ}$ lie within the error range of the experimental data. The $\eta_2(1870)$ has been observed in the invariant mass spectra of $\eta \pi \pi$, $\pi a_2(1320)$, $f_2(1270) \pi$, and $a_0(980)\pi$ channels. At $\theta = 30^{\circ}$, the calculated ratio $\Gamma(\pi a_2(1320)) / \Gamma(\eta f_2(1270)) = 6.07$ is in close agreement with the experimental value of $4.1 \pm 2.3$ reported by the CBAR Collaboration in 1996 \cite{WA102:1997gkz}. Additionally, the ratio $\Gamma(\pi a_2(1320)) / \Gamma(\pi a_0(980)) = 18.74$ is consistent with the experimental value of $32.6 \pm 12.6$ \cite{WA102:1999ybu} within the uncertainty range. Finally, the calculated ratio $\Gamma(\pi a_0(980)) / \Gamma(\eta f_2(1270)) = 0.32$ is in good agreement with the experimental value $0.48 \pm 0.45$ \cite{Anisovich:2010nh} (see Fig. \ref{fig1} for more details).

\begin{table*}[htbp]
\centering
\caption{The two-body OZI-allowed decay behavior of the $\eta_2(1D)$ and $\eta^{\prime}_2(1D)$ states. All results are in units of MeV. \label{1d}}
\renewcommand\arraystretch{1.5}
\begin{tabular*}{103mm}{@{\extracolsep{\fill}}lcccccccccr}
\toprule[1.00pt]
\toprule[1.00pt]
State && $\eta_2(1D)$ &&&&&&& $\eta^{\prime}_2(1D)$\\
\cline{1-1} \cline{2-4} \cline{9-11} 
Decay channel  &$\theta=0^{\circ}$  &$\theta=15^{\circ}$ &$\theta=30^{\circ}$   &&& &&$\theta=0^{\circ}$  &$\theta=15^{\circ}$ &$\theta=30^{\circ}$   \\
\cline{2-11}
$a_2(1320)\pi$  &173.3    & 161.8  &130.0   && &&& $\backslash$ & 15.3   &57.2\\
$\rho\rho$       &28.5       & 26.6    &21.4&&&&&  $\backslash$     & 8.3    & 31.0    \\
$\omega\omega$   &7.7       &  7.2  &5.8  &&&&&    $\backslash$    &  2.7 &  10.0\\
$a_0(980)\pi$   & 7.0     &  6.5  &5.2&&&&&   $\backslash$  &  0.8  & 3.1\\
$a_1(1260)\pi$   & 4.2     &  3.9 &3.2 &&&&&    $\backslash$  & 2.4  & 8.9\\
$ K{K^*}$         &2.6       &  10.1 &20.7&&&&  & 79.8 &60.4 &39.7  \\
$ K^*{K^*}$    & $\backslash$ & $\backslash$ & $\backslash$   &&&&      & 29.5  &32.8  &30.0 \\
$\eta f_2(1270)$   & $\backslash$ & $\backslash$ & $\backslash$  &&&& & 0.2 & 1.8  & 9.4\\
\midrule[0.7pt]
Total width& 223.3  &216.1  & 186.3&&&& & 109.5  & 123.5 &189.3 \\
\midrule[0.7pt]
Exp   & \multicolumn{3}{c}{$180^{+40}_{-21}\pm25$  \cite{CrystalBarrel:1996bnu}}&&&&&  \multicolumn{3}{c}{$221\pm82\pm44$ \cite{CrystalBall:1991zkb}}\\
\bottomrule[1.00pt]
\bottomrule[1.00pt]
\end{tabular*}
\end{table*}

\subsubsection{The $n=2$ case}

The strong decay behaviors of the $2D$ states are presented in Table \ref{2d}. The experimental decay width of the $\eta_2(2D)$ state is $205 \pm 10 \pm 15$ MeV, which is in good agreement with our calculated results for mixing angles of $\theta = 0^\circ$ and $\theta = 15^\circ$, indicating a small mixing angle. Assuming the mixing angle lies between $0^\circ$ and $15^\circ$, we find that the primary decay channels for the $\eta_2(2D)$ are $\pi a_2(1700)$ and $\pi a_2(1320)$, with decay widths ranging from 91.7 MeV to 85.5 MeV and 51.7 MeV to 48.3 MeV, respectively. The branching ratios for these channels are approximately 43\% and 24\%. Additionally, the $\rho b_1(1235)$ and $\rho \rho$ channels show significant contributions, with decay widths ranging from 20.6 MeV to 19.2 MeV for $\rho b_1(1235)$ and from 19.2 MeV to 17.9 MeV for $\rho \rho$, each contributing about 10\% to the total decay width.

The branching ratios for $\eta_2(2D)$ are reported in Refs. \cite{Anisovich:2000mv, Anisovich:2010nh}, with the updated values in Ref. \cite{Anisovich:2010nh} given as $\mathcal{BR}(a_0 \pi) / \mathcal{BR}(a_2 \pi) = 0.10 \pm 0.08$ and $\mathcal{BR}(f_2 \eta) / \mathcal{BR}(a_2 \pi) = 0.13 \pm 0.06$. In comparison, our calculations yield $\mathcal{BR}(a_0 \pi) / \mathcal{BR}(a_2 \pi) = 0.064$ and $\mathcal{BR}(f_2 \eta) / \mathcal{BR}(a_2 \pi) = 0.060$ for mixing angles $\theta$ between $0^\circ$ and $15^\circ$. The agreement between our theoretical predictions and experimental results, within the reported uncertainties, underscores the reliability of our study. However, the experimental uncertainties remain considerable. Our calculations suggest that more precise measurements of the decay widths and branching ratios of the $\eta_2(2D)$ state would be essential for improving the accuracy of the theoretical models.


The total decay width of $\eta_2'(2D)$ varies between 136.9 MeV and 148.2 MeV as the mixing angle increases from $0^\circ$ to $15^\circ$. Four primary decay channels dominate: $K K^*(1410)$, $K K^*$, $K K^*_2(1430)$, and $K^* K^*$. Among these, the decay widths for $K K^*_2(1430)$, $K K^*$, and $K^* K^*$ exhibit minimal variation with the mixing angle, while the width of the $K K^*(1410)$ channel shows a notable decrease, ranging from 44.0 MeV at $\theta = 0^\circ$ to 27.3 MeV at $\theta = 15^\circ$. Moreover, at $\theta = 30^\circ$, several new decay channels become significant for $\eta_2'(2D)$, including $\pi a_2(1700)$ (26.4 MeV), $\pi a_2(1320)$ (23.6 MeV), and $\pi b_1(1235)$ (18.4 MeV).

\begin{table*}[htbp]
\centering
\caption{The two-body OZI-allowed decay behavior of the $\eta_2(2D)$ and $\eta^{\prime}_2(2D)$ states. All results are in units of MeV.}\label{2d}
\renewcommand\arraystretch{1.5}
\begin{tabular*}{103mm}{@{\extracolsep{\fill}}lcccccccccr}
\toprule[1.00pt]
\toprule[1.00pt]
State && $\eta_2(2D)$ &&&&&&& $\eta^{\prime}_2(2D)$\\
\cline{1-1} \cline{2-4} \cline{9-11} 
Decay channel  &$\theta=0^{\circ}$  &$\theta=15^{\circ}$ &$\theta=30^{\circ}$   &&& &&$\theta=0^{\circ}$  &$\theta=15^{\circ}$ &$\theta=30^{\circ}$   \\
\cline{2-11}
$\pi a_2(1700)$   & 91.7   & 85.5 &68.7 &&&& & $\backslash$ & 7.1  & 26.4\\
$\pi a_2(1320)$  & 51.7  & 48.3 & 38.8 &&&& & $\backslash$  & 6.3 &23.6\\
$\rho b_1(1235)$    & 20.6    & 19.2  & 15.4 &&&& & $\backslash$ & 4.9 &18.4\\
$\rho\rho$             & 19.2  & 17.9& 14.4&&&& & $\backslash$ & 3.0  & 10.1\\
$\omega h_1(1170)$      & 6.3    &5.9  & 4.8 &&&& & $\backslash$  & 2.3  &8.5\\
$\omega\omega$        & 6.0    &5.6  & 4.5 &&&& & $\backslash$  & 1.0 & 3.6\\
$\pi a_0(1450)$       & 4.9    &4.6  & 3.7 &&&& & $\backslash$  & 0.5  & 1.9\\
$a_0(980)\pi$        &3.3   & 3.1   & 2.5 &&&& &$\backslash$&$9 \times 10^{-4}$&$3 \times 10^{-3}$\\
$\eta f_2(1270)$         & 3.1  & 3.0  & 2.5 &&&& & $2\times 10^{-2}$ & 0.1  & 2.9\\
$a_1(1260)\pi$           &2.2    &2.1 &1.6 &&&& &$\backslash$&$3 \times 10
^{-2}$&0.1\\
$ K{K^*}(1410)$      & 1.3   & 4.8  &9.6 &&&& & 44.0  &27.3  & 17.2\\
$ KK^*$            &   0.4 & 2.7  & 6.7 &&&& & 23.4  &24.1  & 17.7\\
$K^*{K^*}$ &0.2&0.3&0.8&&&&   & 17.9 & 18.0 & 15.3\\
$K{K_2^*}(1430)$&0.1&0.3&0.8 &&&& & 35.7 & 39.8  & 35.9\\
$K{K_0^*}(1430)$&$4\times 10^{-2}$&$6\times 10^{-3}$&0.1 &&&& & 7.0  &6.0 &5.7\\
$K_1{K^*}$ &$\backslash$&$\backslash$&$\backslash$ &&&& & 8.9  & 7.2 & 6.4\\
$\omega \omega(1420)$&$\backslash$&$\backslash$&$\backslash$ &&&& & $\backslash$  & 0.6 & 2.2\\
\midrule[0.75pt]
Total width& 211.0  & 203.3& 174.9 &&&& &136.9&148.2&195.9\\ 
\midrule[0.75pt]
Exp   &  \multicolumn{3}{c}{$205\pm10\pm15$ \cite{Anisovich:2000mv}} &&&&&  \multicolumn{3}{c}{$\backslash$}\\
\bottomrule[1.00pt]
\bottomrule[1.00pt]
\end{tabular*}
\end{table*}

\subsubsection{The $n=3$ case}

Data from the reaction $p \bar{p} \to \eta^{\prime} \pi^0 \pi^0$ reveal a $2^{-+}$ signal with a mass of $M = 2267 \pm 14$ MeV and a width of $\Gamma = 290 \pm 50$ MeV, corresponding to the $\eta_2(2250)$ \cite{Anisovich:2000ut}. The $\eta_2(2250)$ is currently listed as a tentative state in the PDG \cite{ParticleDataGroup:2024cfk}, but the available experimental data is insufficient to resolve discrepancies in its characterization. Assuming that the $\eta_2(2250)$ corresponds to the $3^1D_2$ state, we analyze its decay behavior, as shown in Table \ref{3d}. However, our calculations do not reproduce the experimental width of $290 \pm 50$ MeV for any of the mixing angles ($\theta = 0^\circ, 15^\circ, 30^\circ$) \cite{Anisovich:2000ut}, yielding significantly smaller decay widths. Moreover, the total decay width shows minimal sensitivity to variations in the mixing angle.

The dominant decay channel for the $\eta_2(3D)$ state is $\pi a_2(1700)$, which remains the largest contributor across all mixing angles, suggesting its primary role in the total decay width. Additional channels such as $\pi a_2(1320)$ and $\rho \rho$ also contribute notably, though their contributions decrease slightly as the mixing angle increases. Smaller but non-negligible contributions come from channels like $\pi a_0(980)$, $\rho b_1(1235)$, and $\omega \omega$.

For the $\eta^{\prime}_2(3D)$ state, the decay pattern shifts significantly, with channels like $K K^*$ and $K K^*_2(1430)$ emerging as prominent contributors, particularly at higher mixing angles. These channels significantly increase the total decay width of the $\eta^{\prime}_2(3D)$. Since no experimental data is available for the $\eta^{\prime}_2(3D)$, these values remain theoretical predictions. Further studies or observations are needed to confirm its decay properties.

In summary, the total decay width of the $\eta_2(3D)$ state shows a slight decrease with increasing mixing angle, while the decay width of the $\eta^{\prime}_2(3D)$ state increases. This reflects the distinct decay dynamics between the $\eta_2$ and $\eta_2'$ families in their third radial excitation states.

\begin{table*}[htbp]
\centering
\caption{The two-body OZI-allowed decay behavior of the $\eta_2(3D)$ and $\eta^{\prime}_2(3D)$ states. All results are in units of MeV.}\label{3d}
\renewcommand\arraystretch{1.5}
\begin{tabular*}{103mm}{@{\extracolsep{\fill}}lcccccccccr}
\toprule[1.00pt]
\toprule[1.00pt]
State && $\eta_2(3D)$ &&&&&&& $\eta^{\prime}_2(3D)$\\
\cline{1-1} \cline{2-4} \cline{9-11} 
Decay channel  &$\theta=0^{\circ}$  &$\theta=15^{\circ}$ &$\theta=30^{\circ}$   &&& &&$\theta=0^{\circ}$  &$\theta=15^{\circ}$ &$\theta=30^{\circ}$   \\
\cline{2-11}
$\pi a_2(1700)$           & 29.5  & 27.6 & 22.2 &&&& &$\backslash$ &3.7&13.9\\
$\pi a_2(1320)$           & 18.4 & 17.1 & 13.8 &&&& &$\backslash$ &3.7&13.8\\
$\rho\rho$                &10.7 & 10.0  & 8.0 &&&& &$\backslash$&2.9&10.9\\
$\pi a_0(980)$            & 6.1 & 5.7 &4.6 &&&& &$\backslash$&0.1&0.5\\
$\rho b_1(1235)$          & 6.1 &5.7  & 4.6 &&&& &$\backslash$&3.1&11.7\\
$\pi a_1(1260)$           & 3.8 &3.5 &2.8 &&&& &$\backslash$&0.4&1.6\\
$\omega\omega$            &3.3 & 3.1 & 2.5 &&&& &$\backslash$&1.0&3.5\\
$\rho\rho(1450)$          &2.5 & 2.3 &1.9 &&&& &$\backslash$&1.9&7.2\\
$\omega h_1(1170)$        &2.4 & 2.2 & 1.8 &&&& &$\backslash$&1.4&5.1\\
$\omega\omega(1420)$      &1.6 & 1.5 &1.2 &&&& &$\backslash$&0.7&2.7\\
$\eta f_0(1500)$          & 1.6 & 1.5 &1.2 &&&& & $\backslash$&$1\times 10^{-2}$&$4 \times 10^{-2}$\\
$\eta f_2(1270)$      & 1.3 & 1.2 &1.1 &&&& &$3\times 10^{-2}$&0.3&1.6\\
$\pi a_0(1450)$           & 1.3 & 1.2 &0.9 &&&& &$\backslash$&0.4&1.4\\
$K^* K_1$&0.7&0.4&0.2 &&&&  & 5.6 &12.2 & 10.2\\
$K{K^*}(1410)$        & 0.2& 1.6 & 3.9 &&&& &13.5&19.7&15.0\\
$K{K^*}$              &0.1 & 1.1 & 3.1 &&&&  &16.0  &  14.1  & 10.6\\
$K{K_2^*}(1430)$      & 0.1 &0.4 & 1.2 &&&&  & 20.3 & 26.5 &22.7\\
$K{K_0^*}(1430)$&$1\times 10^{-2}$&$3\times 10^{-2}$&$0.2$ &&&& & 3.5 & 4.6 &4.2\\
$K K_1(1650)$&$1\times 10^{-2}$&$7\times 10^{-3}$&$4\times 10^{-3}$ &&&& &0.2&1.1&0.8\\
$K^* K^*(1410)$&$\backslash$&$\backslash$&$\backslash$ &&&& &18.9&16.8&13.9\\
$K^*K^{\prime}_1$&$\backslash$&$\backslash$&$\backslash$ &&&& &5.5&5.0&4.0\\
$K^* K(1460)$&$\backslash$&$\backslash$&$\backslash$ &&&& &4.0&5.6&4.3\\
$\pi(1300) a_0(980)$&$\backslash$&$\backslash$&$\backslash$ &&&& &$\backslash$ &0.5&1.9\\
$a_1(1260) a_1(1260)$&$\backslash$&$\backslash$&$\backslash$ &&&& &$\backslash$&0.5&2.0\\
\midrule[0.75pt]
Total width&89.9 &84.9 & 76.0 &&&& &101.8 &142.9 &177.6\\ 
\midrule[0.75pt]
Exp   &  \multicolumn{3}{c}{$290\pm50$ \cite{Anisovich:2000ut}}&&&&&  \multicolumn{3}{c}{$\backslash$}\\
\bottomrule[1.00pt]
\bottomrule[1.00pt]
\end{tabular*}
\end{table*}

\subsubsection{The $n=4$ case}

Table \ref{4d} presents the two-body OZI-allowed decay widths for the $\eta_2(4D)$ and $\eta^{\prime}_2(4D)$ states across three mixing angles ($\theta = 0^\circ, 15^\circ, 30^\circ$).

For the $\eta_2(4D)$ state, the dominant decay channels are $\pi a_2(1700)$, $\pi a_2(1320)$, and $\rho \rho(1450)$, particularly at lower mixing angles. The total decay width decreases slightly with increasing mixing angle, from 57.9 MeV at $\theta = 0^\circ$ to 52.0 MeV at $\theta = 30^\circ$, indicating minimal sensitivity to the mixing angle. Most decay channels contribute less than 5 MeV, suggesting a relatively spread-out decay distribution without a single, overwhelmingly dominant channel. Our results indicate that the $\eta_2(4D)$ state is relatively narrow. These decay predictions provide essential insights to guide future experimental searches.

In contrast, the $\eta^{\prime}_2(4D)$ state exhibits distinct decay behavior, with channels such as $K K^*$, $K K^*_2(1430)$, and $K^* K^*$ becoming increasingly prominent. These channels contribute significantly to the total decay width. The total decay width for $\eta^{\prime}_2(4D)$ increases notably, from 98.5 MeV at $\theta = 0^\circ$ to 171.0 MeV at $\theta = 30^\circ$, indicating a stronger sensitivity to the mixing angle compared to the $\eta_2(4D)$ state. The $\eta^{\prime}_2(4D)$ state also shows a higher preference for strange decay channels, consistent with its quark content and the effect of mixing.

When compared to the experimental width of $\Gamma = 200 \pm 8^{+20}_{-17}$ MeV for the $X(2600)$ \cite{BESIIICollaboration:2022kwh}, our calculated decay width for the $\eta^{\prime}_2(4D)$ is significantly lower, remaining well below the measured value. This discrepancy rules out the possibility of identifying the $X(2600)$ as an $\eta^{\prime}2(4D)$ state. The $X(2600)$, observed decaying to $\eta' f_0(1500)$ with a combined branching ratio of $\mathcal{B}(J/\psi \to \gamma X(2600)) \times \mathcal{B}(X(2600) \to f_0(1500) \eta') \times \mathcal{B}(f_0(1500) \to \pi^+ \pi^-) = (3.39 \pm 0.18{-0.66}^{+0.91}) \times 10^{-5}$ \cite{BESIIICollaboration:2022kwh}, shows decay patterns that are inconsistent with those expected for the $\eta^{\prime}_2(4D)$ state. Additionally, the OZI rule predicts suppression of the decay of $X(2600)$ into $\eta' f_0(1500)$, further supporting the exclusion of the $\eta^{\prime}_2(4D)$ identification.  Given the substantial experimental uncertainties and the limitations of current theoretical models, we emphasize the need for more precise measurements of the total decay width and decay patterns of the $X(2600)$ to validate these conclusions.

Overall, the $\eta_2(4D)$ state exhibits a more distributed decay pattern with minimal sensitivity to the mixing angle, while the $\eta^{\prime}_2(4D)$ state shows a clear shift towards kaon-related decays and a stronger dependence on the mixing angle. These differences highlight the need for further exploration of mixing angles and additional decay modes to fully understand the decay characteristics of these high-excitation $\eta_2$ states.

\begin{table*}[htbp]
\centering
\caption{The two-body OZI-allowed decay behavior of the $\eta_2(4D)$ and $\eta^{\prime}_2(4D)$ states.  All results are in units of MeV.}\label{4d}
\renewcommand\arraystretch{1.5}
\begin{tabular*}{103mm}{@{\extracolsep{\fill}}lcccccccccr}
\toprule[1.00pt]
\toprule[1.00pt]
State && $\eta_2(4D)$ &&&&&&& $\eta^{\prime}_2(4D)$\\
\cline{1-1} \cline{2-4} \cline{9-11} 
Decay channel  &$\theta=0^{\circ}$  &$\theta=15^{\circ}$ &$\theta=30^{\circ}$   &&& &&$\theta=0^{\circ}$  &$\theta=15^{\circ}$ &$\theta=30^{\circ}$   \\
\cline{2-11}
$\pi a_2(1700)$           & 18.0&16.8&13.5&&&&     &$\backslash$&2.8&10.3\\
$\pi a_2(1320)$           &9.9&9.2&7.4&&&& &$\backslash$&2.6&9.5\\
$\rho\rho(1450)$          & 5.2&4.8&3.9&&&& &$\backslash$&2.8&10.4 \\
$\pi a_0(980)$           & 4.7&4.4&3.5&&&& &$\backslash$&0.3&1.1\\
$\rho\rho$                & 4.1&3.9&3.1&&&&  &$\backslash$ &1.8&6.6\\
$\rho b_1(1235)$          & 3.9&3.7&2.9&&&&  &$\backslash$ &2.3&8.5\\
$\pi a_1(1260)$           &2.4&2.2&1.8 &&&& &$\backslash$&0.4&1.6\\
$\omega\omega(1420)$      &2.1&2.0&1.6&&&&  &$\backslash$ &1.0&3.8\\
$\omega h_1(1170)$        &1.7&1.6&1.3&&&&  &$\backslash$&1.0&3.7\\
$\pi(1300) a_0(980)$      &1.3&1.2& 1.0&&&& &$\backslash$&0.2&0.8\\
$\omega\omega$            &1.3&1.2&1.0 &&&&  &$\backslash$&0.6& 2.1\\
$\eta f_0(1500)$      &1.1&1.1&0.8 &&&& &$\backslash$&$1 \times 10^{-2}$&$5 \times 10^{-2}$\\
$\eta f_2(1270)$ &0.8&0.8&0.6 &&&&  &$1\times 10^{-2}$ &0.2&1.1\\
$K_1 K^*$ &0.3&0.3&0.3&&&&   &4.8 &9.8&7.9\\
$a_1(1260)a_0(980)$ &0.3&0.2&0.2 &&&&   &$\backslash$ &0.3&1.3\\
$K{K^*}(1410)$        & 0.2&0.8&2.1 &&&& &6.2 &11.7&9.9\\
$K^* K^*(1410)$&0.2&0.1&0.3 &&&&   &20.7 &13.7&11.6\\
$K{K_2^*}(1430)$      & 0.1&0.4 &1.4 &&&& &15.4&19.4&16.0 \\
$K^*{K_2^*}(1430)$    & 0.1&0.6 &1.3 &&&& &6.1&11.5&9.5\\
$K^* K^*$ &$7 \times 10^{-2}$ &0.3&1.0 &&&&   &11.0 &15.6&12.8\\
$K{K^*}$        &$3\times 10^{-2}$ &0.7 &2.1 &&&& &12.9 &11.6&9.0\\
$K^{\prime}_1 K^*$&$2 \times 10^{-2}$&0.2&0.6 &&&&   &5.1 &10.6&8.6\\
$K^*K(1460)$ &$2\times 10^{-2}$&$6 \times 10^{-2}$&0.1 &&&&  &6.1 &5.7 & 4.4\\
$K K^*_0(1430)$ &$3 \times 10^{-4}$&$4 \times 10^{-2}$&0.2 &&&&    &1.9&2.8&2.4\\
$K^*{K_1}(1650)$ &$\backslash$&$\backslash$&$\backslash$ &&&&   &7.1 &15.8&12.7\\
$K_1 K^*(1410)$  &$\backslash$&$\backslash$&$\backslash$ &&&&    &2.4&0.9&1.2\\
$K_1K^*_2(1430)$ &$\backslash$&$\backslash$&$\backslash$ &&&&   &1.2 &1.3&0.9\\
$a_1(1260)a_2(1320)$ &$\backslash$&$\backslash$&$\backslash$ &&&&   &$\backslash$&0.5&1.7 \\
$\rho(1450) b_1(1235)$ &$\backslash$&$\backslash$&$\backslash$ &&&&   &$\backslash$&0.4&1.6\\
\midrule[0.75pt]
Total width&57.9&55.9&52.0 &&&& & 98.5&147.6&171.0 \\ 
\bottomrule[1.00pt]
\bottomrule[1.00pt]
\end{tabular*}
\end{table*}

\section{summary}\label{sec3}

Inspired by the discovery of the $X(2600)$, this study explores isoscalar $2^{-+}$ mesonic states, with a focus on the $\eta_2$ meson family. Using a modified GI model, we calculate the mass spectrum for $\eta_2$ states, ranging from $\eta_2(1D)$ to $\eta_2(4D)$. Our theoretical results are compared with experimental data, showing good agreement with the observed masses. Additionally, we investigate the OZI-allowed two-body strong decay behaviors by utilizing the wave functions derived from the modified GI model.

Our analysis of the mass spectrum does not support the identification of the $X(2600)$ as the $\eta^{\prime}_2(4D)$ state. To further evaluate this hypothesis, we examine the two-body OZI-allowed strong decays of the $\eta^{\prime}_2(4D)$. Comparing the theoretical decay patterns and widths with the experimental data from BESIII, we observe discrepancies between the predicted and measured decay widths for the $X(2600)$, indicating that it is unlikely to correspond to the $\eta^{\prime}_2(4D)$ state.

This comprehensive study offers valuable insights into the properties of the $\eta_2^{(\prime)}$ mesons, highlighting the complexities of meson spectroscopy and the importance of multidimensional approaches for accurately identifying and classifying new particles. Our analysis covers the $\eta_2(1645)$ and $\eta_2(1870)$ states, as well as their radial excitations. By examining these states collectively, we provide a broader context for understanding the behavior of $\eta_2^{(\prime)}$ mesons and meson spectroscopy in general. This holistic approach contributes to a more complete picture of the $\eta_2^{(\prime)}$ meson family and offers guidance for future experimental and theoretical investigations.

\section*{Acknowledgments}
LMW would like to thank the Lanzhou Center for Theoretical Physics for supporting his stay at Lanzhou University, where this work was completed.
This work is supported by  the National Natural Science Foundation of China under Grants No.~12335001, No.~12247101 and No.~12405104, National Key Research and Development Program of China under Contract No.~2020YFA0406400, the 111 Project under Grant No.~B20063, the fundamental Research Funds for the Central Universities, the project for top-notch innovative talents of Gansu province, Natural Science Foundation of Hebei Province under Grants No.~A2022203026, the Higher Education Science and Technology Program of Hebei Province under Contract No.~BJK2024176, the Research and Cultivation Project of Yanshan University under Contract No.~2023LGQN010.

\end{document}